**Title: Modeling and Measurement of Lead Tip Heating in Implanted Wires with Loops**

Lydia J Bardwell Speltz (1), Seung-Kyun Lee (2), Yunhong Shu (1), Matt A Bernstein (1)

**Author Affiliations and Addresses:**

(1) Department of Radiology, Mayo Clinic, Rochester, MN, United States

Address: Mayo Clinic
200 First St. SW
Rochester, MN 55905

(2) GE HealthCare Technology & Innovation Center, Niskayuna, NY, United States.

Address:
1 Research Circle
Niskayuna, New York 12309

**Corresponding Author Info:**

Name: Yunhong Shu

Address: Mayo Clinic
200 First St. SW
Rochester, MN 55905

Phone number: 507-255-3325

Email: shu.yunhong@mayo.edu

**Purpose:** To theoretically and experimentally study implant lead tip heating caused by radiofrequency (RF) power deposition in different wire configurations that contain loop(s).

**Methods:** Maximum temperature rise caused by RF heating was measured at 1.5T on 20 insulated, capped wires with various loop and straight segment configurations. The experimental results were compared with predictions from the previously reported simple exponential and the adapted transmission line models, as well as with a long-wavelength approximation.

**Results:** Both models effectively predicted the trends in lead tip temperature rise for all the wire configurations, with the adapted transmission line model showing superior accuracy. For superior/inferior (S/I)-oriented wires, increasing the number of loops decreased the overall heating. However, when wires were oriented right/left (R/L) where the *x*-component of the electric field is negligible, additional loops increased the overall heating.

**Conclusion:** The simple exponential and the adapted transmission line models previously developed for, and tested on, straight wires require no additional terms or further modification to account for RF heating in a variety of loop configurations. These results extend the models' usefulness to manage implanted device lead tip heating and provide theoretical insight regarding the role of loops and electrical lengths in managing RF safety of implanted devices.

# 1. Introduction

Radiofrequency (RF) power deposition at the lead tips can cause substantial heating, posing a major safety concern for patients with implanted devices undergoing MRI examinations [1-3]. Many of these devices are specifically designed and tested for scanning at a magnetic field strength of 1.5T, which is commonly used clinically for patients who have implanted devices. Most active implants, including cardiac implantable electronic devices, deep brain stimulators, spinal cord stimulators, etc., contain an implantable pulse generator (IPG) and one or more leads. When leads are surgically implanted, their trajectories often include single or multiple loops of various sizes. Understanding how loops influence lead tip heating remains an ongoing area of research [4-10]. A model that can be applied to predict lead tip heating with various lead configurations could potentially inform lead placement strategies and improve safety protocols, i.e., allow greater access to MRI for patients with implants without increasing the risk of adverse effects.

Several studies investigating the impact of loops on lead tip heating have primarily focused on deep brain stimulation (DBS) leads. Main findings from these studies are: 1) increasing the number of loops tends to *reduce* heating [4,5,7,8], 2) positioning loops near the burr hole decreases heating [4,6,7] and 3) contralateral leads experience more heating than ipsilateral leads [6]. A limited number of studies on pacemaker or epicardial leads have similarly found that: 1) increasing the number of loops also reduces heating [9], and 2) anterior-facing loops generate more heating than inferior-facing loops [10]. Finding 2) is not surprising because the *z*-component of the B1-field is designed to be relatively small.

In a recent paper [11], our group introduced a simple exponential model and an adapted transmission line model for the transfer function to predict lead tip voltage, and experimentally tested the models by measuring lead tip temperature rise in a series of straight, insulated wires. In this paper, we measure lead tip heating for a series of insulated wires that include both a straight section and loops of various size, number, and placement. We then applied the two models from Reference [11] to test whether they still hold in their original form, or whether they require modification to account for the presence of loops. We also considered a long-wavelength approximation (i.e., setting the transfer function equal to 1) that allows derivation of analytical results, which may provide some insight. The objective of this study is to validate and potentially extend the existing models to better predict lead tip heating in more complex lead configurations, specifically containing loops.

**2. Theory**

Lead tip voltage can be calculated using the complex line integral of the incident electric field multiplied by a suitable transfer function $h$ [2]:

$$V = \int_0^d h(l)\, \mathbf{E}(l) \cdot d\mathbf{l} \quad [1]$$

where $l$ is the distance from the lead tip, $d$ is the total length of the lead, $\mathbf{E}$ is the electric field (which can be complex), and bold font indicates a vector. In Reference [11] we considered two forms of the transfer function: the simple exponential model:

$$h(l) = e^{-ikl} \quad [2]$$

and an adapted transmission line model:

$$h(l) \propto \frac{e^{-ikl}\left(1 - \Gamma e^{-2ik(d-l)}\right)}{(1 - \Gamma e^{-2ikd})} \quad [3]$$

where $\Gamma$ is the reflection coefficient. In both Equations (2) and (3), the complex King wavenumber for $k$ is used, as derived from insulated antenna theory [3,11-15].

The insulated wires described in Reference [11] and used in this work had inner and outer radii of the insulator $a = 0.390$ mm and $b = 0.625$ mm, resulting in a calculated King wavenumber of $(6.7454 - 0.6840\, i)$ rad m$^{-1}$ at 1.5T (for more details see Reference [11].) To calculate the temperature rise $\Delta T \propto |V|^2$ from Equation (1) for a set of wires, the simple exponential model requires only one free parameter per set of data for global scaling, while the adapted transmission line model requires a second free parameter per set of data to fit the reflection coefficient (unless these parameters are obtained by other means, such as modeling.) Both models predict the resonant length without the use of any free parameters. Figure 1 shows plots of the magnitude and phase of the transfer function for the simple exponential model and the adapted transmission line model for various values of length $d$, and the specific value of the wavenumber $k$ we considered.

To account for loops in Equation (1), one approach is to apply the well-known Stokes' theorem [16], which states the line integral of a function over a closed loop defining a bounded surface is equal to the surface integral of the curl of the function. This theorem can be applied in conjunction with one of Maxwell's equations [17]

$$\nabla \times \mathbf{E} = -\frac{d\mathbf{B}}{dt} \qquad [4]$$

to yield Faraday's law of induction. If the magnetic field has sinusoidal time dependence represented by a complex exponential, then the right-hand side of Equation (4) becomes $-i\omega\mathbf{B}$, further simplifying the calculation.

In this paper, we consider the forms of the transfer function given by Equations (2) and (3), as well as their long-wavelength approximation $h = 1$, as described in the Appendix. Even though the simple exponential model of Equation (3) has no additional free parameters compared to the approximation $h = 1$, the latter may be worthwhile to consider because the King wavelength has the relatively large value of 93.1 cm in our case. Notably, the assumptions of $h = 1$ and a uniform B1 field permits an analytical derivation of the relative contributions to lead tip voltage from a straight wire and loops of the same total wire length (see Appendix). That result shows that the voltage contribution from the straight section tends to dominate that of a loop of the same length $d$, independent of the main field and B1 strength. This is generally true except when the straight section is oriented along the midline where the electric field is negligible. Additionally, the derivation in the Appendix demonstrates that the relative contribution from a loop decreases linearly with the number of turns in the loop, if the total length of the wire is held constant.

While the results of the approximate analysis developed in Appendix can explain some findings reported in the literature, the simplified analysis fails to explain some other previously reported experimental findings [4,6,7], e.g., the placement of the loop affects the observed temperature rise. Specifically, it was reported consistently that loops placed near the tip (or the burr hole) reduced heating better than loops farther away from the tip. Consequently, we return to the more complete transfer functions in Equations (2) and (3) to examine if they can model those effects.

Findings related to the positioning of the loop can be understood in terms of the concept of "electrical length", which is captured in the transfer function $h(l)$. In this interpretation, a wire with loops positioned closer to the lead tip are expected to heat up less than a wire with equivalent loops farther away from the tip. This is because the former loops add path length, which attenuates the transfer function (see Figure 1) over a greater proportion of the non-looped path of the line integral.

We hypothesize that Equation (1), in conjunction with a suitable transfer function, can provide quantitative predictions related to three of the findings reported in the literature on the role of loops in lead tip heating: (i) loops reduce heating in general, (ii) more turns, while keeping the total lead length fixed, also reduces heating, and (iii) loops closer to the lead tip tend to result in reduced heating. Through a set of controlled wire heating experiments, we test how well such model predictions agree with measured data, providing insight into the most suitable choice of transfer function.

## 3. Methods

### 3.1 American Society for Testing and Materials (ASTM) Phantom

Temperature rise measurements were obtained using the same methods detailed in Reference [11] and are only briefly reviewed here. The same ASTM phantom was used to house the different configurations of the same type of insulated wire. The "IPG" end of the wire was insulated/capped as described in Reference [11], corresponding to the inferior end of each S/I (i.e., "vertically-") oriented wires and the patient-left end of the R/L (i.e., "horizontally-") oriented wires. The "lead tip" end of the wire had 5 mm of the insulation removed, corresponding to the superior end of each vertically-oriented wire and patient-right end of each horizontally-oriented wire. The phantom followed the ASTM F2182-11a standard [18] and the gel had a relative permittivity of 80 and a conductivity of 0.47 S/m. Coronal views of the wire configurations are shown schematically in Figures 2 and 3 for vertically- and horizontally-oriented straight segments, respectively. The vertical wire paths in Figure 2 are superimposed on the magnitude of the square root of the sum of squares of the $E_x, E_y$ and $E_z$ components of the calculated electric field, and the horizontal paths are superimposed on the magnitude of the $E_x$ component as shown in Figure 3. More details about the electric field calculation are provided in Reference [11]; particularly related to Figure 4a of that paper. Each wire configuration either included zero loops (i.e., a straight wire), one loop, or two loops. The S/I center of each wire configuration was placed at isocenter as described in Methods in Reference [11]. All loops and straight sections were coplanar on a mid-line coronal plane of the scanner ($y = 0$). The loop diameters are provided in Table 1, with most loops approximately 5 cm in diameter, and larger loops approximately 7 cm.

### 3.2 Temperature Measurements

Using a 1.5T scanner (HDx 16.0, GE HealthCare, Chicago, IL, USA), temperature measurements were taken at the lead tips for all 20 different wire configurations. As in Reference [11], temperature measurements were collected using a Fluoroptic® thermometer (Luxtron model FOT Lab Kit, Lumasense Technologies, Santa Clara, CA, USA) with a resolution of 0.1°C and sampling rate of 10 $s^{-1}$. The same TG- correction that was described in Methods in Reference [11] was applied to these temperature measurements, with the purpose of equalizing the applied B1+ values across the set of wires.

### 3.3 Simulated, Spatially Varying Electric Field

The simulated lead tip voltages were calculated using a spatially varying, simulated electric field **E**, along with Equation (1) and the transfer function from either Equation (2) for the simple exponential model or Equation (3) for the adapted transmission line model. The electric field generated by the 1.5T body transmit coil were simulated using Sim4Life (ZMT, Zurich, Switzerland) as described in References [11, 19]. For reference, we also repeated the calculation with the long-wavelength approximation, $h = 1$, but note that this calculation used the simulated electric field, which is more sophisticated than the simplifying assumption of a constant electric field made in the Appendix. In all cases, the temperature rise was then calculated using the relationship $\Delta T \propto |V|^2$.

### 3.4 Statistical Analysis

We identified three sources of error in our temperature measurements, as described in our previous work [11]. The first source of error arises from the variation in temperature rise recorded by the two different fluoroptic probes positioned at the lead tip. The second source of error stems from any fluctuation in transmit gain (TG) values calibrated during prescan. The third source of error is the standard deviation of the baseline temperature recorded before any RF power was applied. The overall error, represented by the error bars in Figures 4-6, was calculated by taking the square root of the sum of the squares of these three types of error.

A single scaling factor was fitted to the temperature rise data for each transfer function model prediction in order to minimize root mean square error (RMSE) over the entire set of 20 wires. As described in Theory, the adapted transmission line model, includes an additional parameter for the reflection coefficient. Since absolute truth data was not available, we used the Akaike Information Criterion (AIC) to compare the models, accounting for the different number of free parameters. The AIC was calculated using ordinary least squares, as outlined in [20].

## 4. Results

Figure 4 compares the predictions of the simple exponential model and the adapted transmission line model against the measurements for the 20 different wires. The fitted reflection coefficient $\Gamma$ that appears in Equation (3) was determined to be 0.2403, reasonably close to the value of 0.2759 fitted from the straight wire data at 1.5T in Reference [11]. Figure 7 shows a plot of the temperature rise versus the experimental data for the $h = 1$ approximation.

To better understand the results, we plot some of the data in Figure 4 separately in individual subplots, especially because the maximum temperature rise has a large dynamic range across the experiments. To assess the influence of overall wire length on heating, we conducted group comparisons that controlled for either the total height (i.e., S/I extent of the wire configuration, regardless of any loop(s)) of the wire $\Delta z$), or the total wire length $d$. The results of these group comparisons are presented in Figure 5, which includes: (A) vertically-oriented wires with the same net height of 25 cm, (B) vertically-oriented wires with total length of 25 cm, (C) vertically-oriented wires with total length of 40.7 cm, and (D) vertically-oriented wires with total length of 47 cm.

Table 2 provides a comparison of the root mean square error (RMSE) and the Akaike Information Criterion (AIC) for the simple exponential model, the adapted transmission line model and the $h = 1$ approximation. Among these, the adapted transmission line model with the optimized $\Gamma$ provided a better fit, achieving the lowest RMSE and AIC values, indicating better predictive accuracy even with the additional free parameter, again consistent with the result of Reference [11]. As expected, the simple exponential model outperformed the $h = 1$ approximation, with no additional free parameters.

In Figure 6, the temperature rise results of the simple exponential model, the adapted transmission line model, and the corresponding experimental data are shown for all the horizontally-oriented wires. Note that the error bars appear relatively large because the measured temperature rise was small (5°C or less) compared to the vertically-oriented wires. Hence, the vertical axis of the plots in Figure 6 has a much smaller range compared to those in Figure 5. Comparing the three cases with the same wire length (horizontally-oriented (H) Straight, H Single Short, and H Double Short whose configurations are shown in Figure 3), it is apparent that adding loops increased the temperature for the horizonal wires.

## 5. Discussion

As illustrated by Figure 4 and Table 2, both the simple exponential model and the adapted transmission line model appear to effectively account for the different loop configurations well, with the adapted transmission line model providing a better quantitative fit. Notably, neither model requires modification or additional terms to Equation (1) to account for loops. Perhaps this can best be understood in light of Equation (4). Because the simulated electric field has a non-zero curl, the magnetic field is implicitly accounted for and is integrated within the model, obviating the need for the magnetic field to be considered separately. The result shows the robustness of these models in handling more complex wire geometries beyond straight wires, especially those involving loops.

A trend that is apparent in much of Figure 4 and throughout Figure 5 is that the introduction of loops and increasing their number of turns $N$ generally *reduce* heating (with some exceptions discussed in the next paragraph) consistent with the findings of previous studies [4,5,7,8,9]. This trend can be understood based on the simplified calculation provided in Appendix, including the scaling relation with the number of loop turns $N$, and aligns well with the concept of electrical length. That is, the portion of the wire used for the loops contributes a relatively small voltage of its own compared to a straight section of the same length, while attenuating the transfer function (see Figure 1) farther along the path of the line integral in Equation (1). This attenuation becomes more pronounced the larger the negative imaginary part of the complex wavenumber (see Reference [11] for further discussion of this sign of Im($k$)).

This reduction in heating with loop(s) is not universal, so caution should be exercised in applying this rule in practice. Note most of the wires shown in Figure 4 and all wires in Figure 5 were positioned vertically. However, guided by the underlying theory, we designed an experiment in which heating was expected to increase with the introduction of loops. The wires shown in Figure 3 were positioned horizontally, where $E_x \approx 0$. As a result, the line integral of the straight segments contribute negligible voltage, so heating is expected to *increase* with the introduction of loops. This theoretical prediction and experimental results are illustrated in Figure 6. Furthermore, as additional loops of the same size were added, the heating effect continues to intensify. The contrasting behavior between vertical and horizontal wires underscores the need for careful consideration of wire geometry when designing and evaluating lead heating for implanted medical devices.

For the vertically-oriented wires, in addition to the number of the loops, the location of the loop also effects overall heating. Wires with loops positioned further from the lead tip (bottom loops) heat more than wires with equivalent loops near the lead tip (top loops). Again, this can be explained with the electrical length concept, because the top loops add path length and attenuate the transfer function over the entire remaining (upstream) straight section of the wire, while the bottom loops do not affect the electrical distance to the lead tip for the preceding (downstream) straight portion of the wire. Given the generally decaying nature of $|h(l)|$ apparent on Figure 1, lead-tip heating is expected to decrease when a loop is located closer to the lead tip (on the superior side of the phantom in these experiments).

Considering the magnitude $|h(l)|$ alone does not explain the result that loops in the middle of the wire achieved the lowest temperature in two, wire length-controlled comparisons shown in Figure 5. This was observed among the first three cases (Bottom, Middle, Top Single; 40.7 cm wire length) and the second set of three cases (Bottom, Middle, Top Double; 56.4 cm wire length) in Figure 5(A), both in experiment and simulation. One possible explanation can come from the phase of the transfer function $\sphericalangle h(l)$, also plotted in Figure 1b. In the mentioned cases, the wire lengths were comparable (within 21%) to $\lambda/2 = 46.6$ cm. This implies that the contributions to the complex path integral Equation (1) from the top and the bottom regions of the wire would have roughly opposed phases, and tend to cancel partially. Inserting a loop in the middle of the wire suppresses the contribution of that section of the wire as explained above, leaving the integral dominated by the two end regions, leading to reduction in the summed voltage. While more experiments are needed for further verification, this consideration may offer a way to suppress resonant heating when the wire length approaches half wavelength.

There are several limitations to this study. The measurements were limited to straight and circular loop segments in insulated wires that were not actual device leads, with one end capped with an insulator. The physical and electrical properties of the wires can differ from those of clinical leads. While this setup serves as a reasonable approximation for capped, abandoned leads, it does not fully represent various clinical scenarios, such as leads connected to IPGs, leads capped with materials other than insulators [21], or helical leads [22]. Additionally, only a limited number of configurations of loops (e.g., all circular) were investigated here. Due to geometrical limitation of the phantom, we only studied loops aligned with the coronal plane of the scanner bore. More complex lead geometries, including those with bends or irregular shapes, were not considered and might exhibit different heating characteristics. Lastly, the study's findings are specific to the 1.5T MRI environment. Different field strengths could result in different heating patterns, especially when resonant lengths match the wire length.

The limitation of our experimental setup such as the use of simplified wire and limited phantom configuration, suggest that further research is necessary. Our future work could include replicating these experiments at 3T and 7T, systematically studying the effect loops have on resonant length, exploring more complex wire geometries, and testing in vivo scenarios to validate these findings and refine our predictive models. Because $h = 1$ is a long-wavelength approximation, we expect that both the simple exponential model and the adapted transmission line model will be considerably more important at field strengths above 1.5T, consistent with results reported in Reference [11].

## 6. Conclusions

This study investigated the maximum temperature rise at the lead tips of various insulated wires placed in a phantom setup at 1.5T, considering different orientation, loop placements and numbers of loops. We compared our experimental results with predictions from the simple exponential model, the adapted transmission line model, and the long wavelength approximation of $h = 1$. Overall, all models qualitatively captured trends in lead tip temperature rise. Quantitatively, the adapted transmission line model provided the most accurate predictions, as evidenced by the lowest Akaike Information Criterion (AIC) values, followed by the simple exponential model. The $h = 1$ performed the worst, despite having the same number of free parameters as the simple exponential model. However, the $h = 1$ model has utility because some general features of the experimental results were captured in the calculation (Appendix) providing an analytical expression for the ratio of voltage contributions of a loop to straight wire, under simplified assumptions.

In conclusion, this study provides insight into how wire looping affects lead tip RF heating and demonstrates potential application of the transfer function-based theoretical models for managing the heating risks.

**Funding:** This work was supported in part by National Institutes of Health U01EB024450.

**Appendix**

Here we examine the relative contributions to lead tip voltage from a straight wire and loops of the same total wire length, under simplified conditions. Consider a uniform B1 field whose magnitude in the y direction is $b1_0$. Because Equation (4) is evaluated in the laboratory (non-rotating) frame, $|dB_y/dt| = \omega\, b1_0$, where $\omega/2\pi$ is the Larmor frequency. For simplicity, let us consider a mid-line coronal plane (i.e., $y = 0$). Then, also from Equation (4), projected along the y axis, the *z*-component of electric field can be written as:

$$E_z = x\,\omega\, b1_0 + c \tag{A1}$$

in a region where $E_z$ dominates $E_x$ (see, for example, Figure 4 of Reference [11]). Here we have omitted the time-dependent sinusoidal factor because it does not affect the line integration in Equation (1). We will set the constant *c* in Equation (A1) equal to zero, which is consistent with the well-known result of a straight wire that is placed midline produces negligible heating. This choice is also consistent with standard EM simulations (e.g., Figure 4 in Reference [11]) showing z-component of electric field is zero in a midline coronal plane when *x*=0.

The transfer functions given by Equation (2), reduces to $h(l) \approx 1$ when $|k|d \ll 1$, which typically occurs when the wire length *d* is much shorter than the King wavelength $\lambda = 2\pi/\text{Re}(k)$. (The same is true for Equation 3, provided that $\Gamma \neq 1$). If we do make the further simplifying assumption that $h(l) = 1$, then for a vertically-oriented (i.e., z-directional), straight wire of length *d* that is displaced R/L from midline by a distance *x*, Equation (1) reduces to

$$V_{\text{straight}} = x\, d\, \omega\, b1_0 \tag{A2}$$

If this straight wire is bent into a circular loop, then the area $A$ of the resulting loop is

$$A = \frac{d^2}{4\pi} \tag{A3}$$

According to Faraday's law of induction, the voltage induced in the loop is $V_{\text{loop}} = A\,\omega\,b1_0$, and combining Equations (A2) and (A3) yields

$$\left|\frac{V_{\text{loop}}}{V_{\text{straight}}}\right| = \frac{d}{4\pi|x|} \tag{A4}$$

Note that the ratio in Equation (A4) is independent of both main magnetic and RF field strength (i.e., $\omega$ and $b1_0$, respectively), and instead only depends on geometrical factors. (Recall, however, that at higher field strengths the long-wavelength approximation $h = 1$ breaks down.) Also note that the straight wire produces greater voltage than the loop whenever $|x| > \frac{d}{4\pi}$, which is usually satisfied unless the straight wire is intentionally placed along the midline.

Equations (A2)-(A4) can be readily generalized to the case of the wire bent into a circular loop with $N$ turns, again holding its total length $d$ constant. Then the area $A$ of the resulting loop becomes

$$A = \frac{d^2}{4\pi N^2} \tag{A5}$$

and the voltage induced in the loop is $V_{\text{loop}} = N\,A\,\omega\,b1_0$, yielding

$$\left|\frac{V_{\text{loop}}}{V_{\text{straight}}}\right| = \frac{d}{4\pi N\,|x|} \tag{A6}$$

From Equation (A6), we can see the inverse linear dependence with the number of turns $N$ in this simplified model. Also note that the condition $|x| > \frac{d}{4\pi N}$ becomes even easier to satisfy as $N$ increases.

Table 1: Total length and total height (Δz) for all wires tested. Note "H" denote horizontal wires (i.e., oriented R/L), while all the rest labels denote vertical wires (i.e., oriented S/I).

|  | Total Length (cm) | Total Height (cm) | # of Loops | Diameter of Loops (cm) |
|---|---|---|---|---|
| Straight | 25 | 25 | 0 | - |
| Bottom Single | 40.7 | 25 | 1 | 5 |
| Middle Single | 40.7 | 25 | 1 | 5 |
| Top Single | 40.7 | 25 | 1 | 5 |
| Bottom Double | 56.4 | 25 | 2 | 5 |
| Middle Double | 56.4 | 25 | 2 | 5 |
| Top Double | 56.4 | 25 | 2 | 5 |
| Straight Long | 40.7 | 40.7 | 0 | - |
| Middle Single Large Loop Long | 47 | 25 | 1 | 7 |
| Middle Single Back | 40.7 | 25 | 1 | 5 |
| Middle Single Flipped | 40.7 | 25 | 1 | 5 |
| Straight Xlong | 47 | 47 | 0 | - |
| Middle Single Large Loop | 40.7 | 18.7 | 1 | 7 |
| Middle Single Short | 25 | 9.3 | 1 | 5 |
| Middle Double Long | 47 | 15.5 | 2 | 5 |
| H Straight | 40 | 40 | 0 | - |
| H Single Short | 40 | 25.5 | 1 | 5.5 |
| H Single Long | 54.8 | 8.2 | 1 | 5.5 |
| H Double Short | 40 | 40 | 2 | 5.5 |
| H Double Long | 72.1 | 40 | 2 | 5.5 |

Table 2: Statistical results for $h(l) = 1$, the simple exponential model (SEM), and adapted transmission line model (TLM) fits. RMSE=root mean square error, AIC= Akaike Information Criterion

|  | h=1 | SEM | TLM |
|---|---|---|---|
| Number of Free Parameters | 1 | 1 | 1 |
| RMSE | 15.52 | 13.35 | 9.95 |
| AIC | 28.89 | 25.87 | 22.00 |

**Figures**

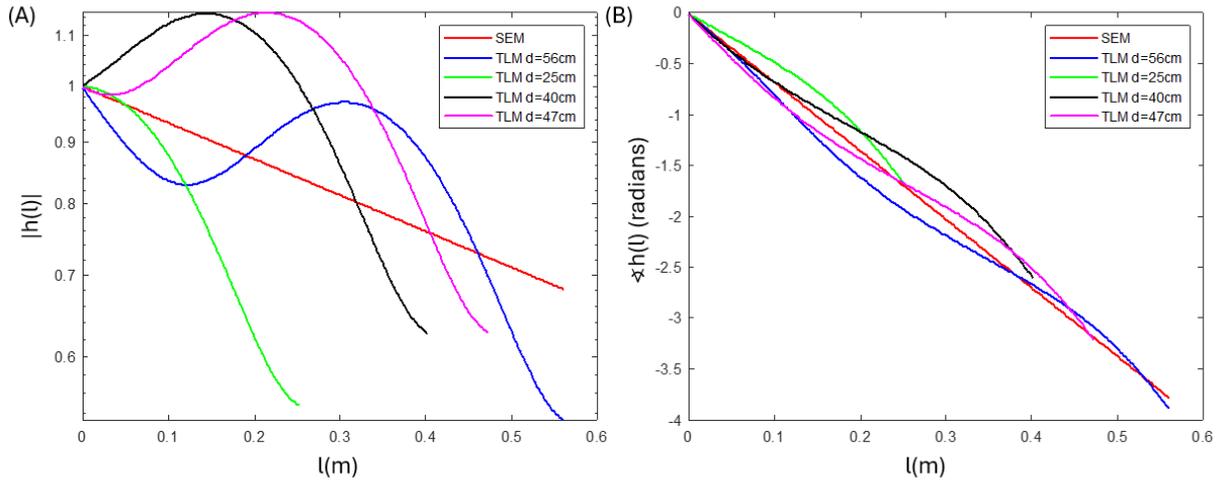

Figure 1: The (A) magnitude and (B) phase of the simple exponential (SEM) and adapted transmission line (TLM) transfer functions $|h(l)|$ plotted versus the length variable $l$ on a log scale. The adapted transmission line transfer function depends on $d$ (total wire length) so the four TLM lines with different $d$ have qualitatively different functional dependencies on $l$. Nevertheless, they all exhibit overall decaying trend with $l$ governed by Im($k$) < 0.

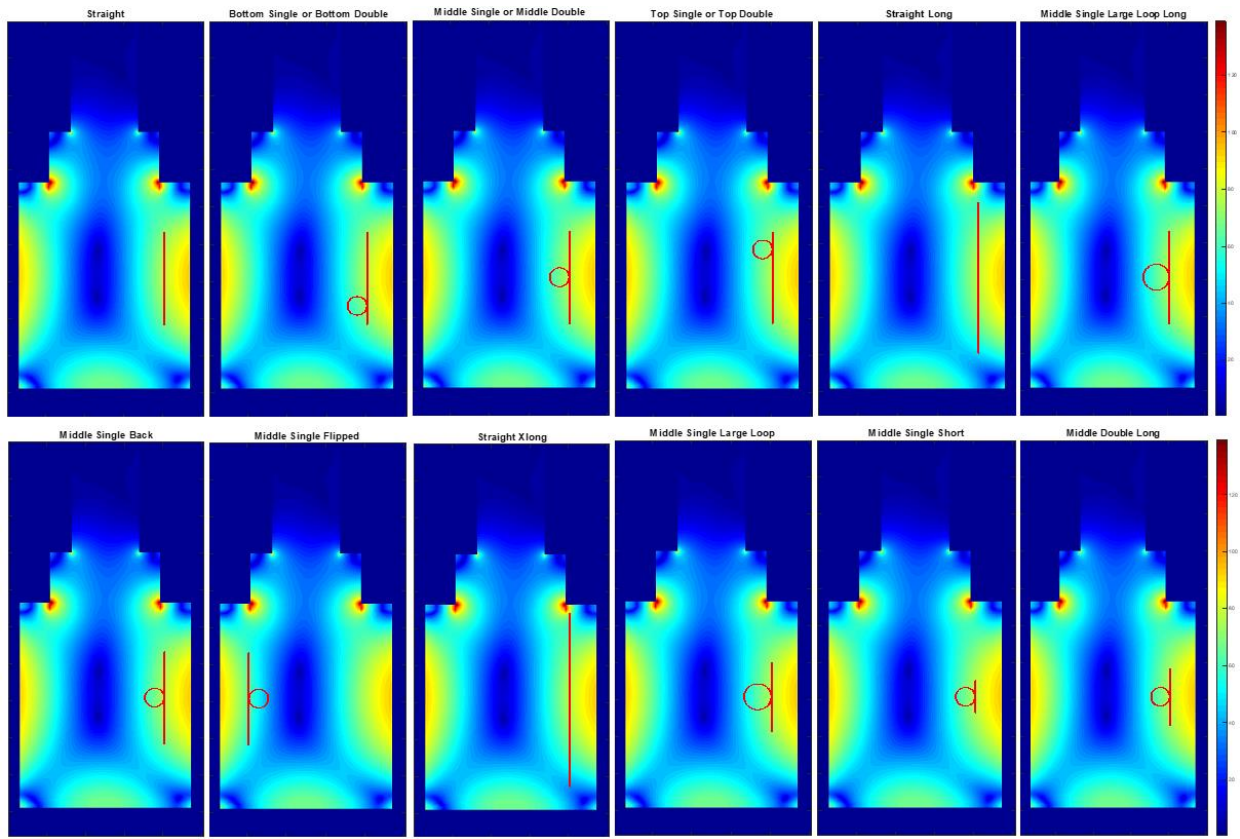

Figure 2: Electric field simulations (see text) for the ASTM phantom at $y = 0$ (i.e., a coronal midline slice) at 1.5T showing all the different configurations of the vertical wires tested. Note that "lead tip" end of the wire corresponds to the superior end of each vertically-oriented wire.

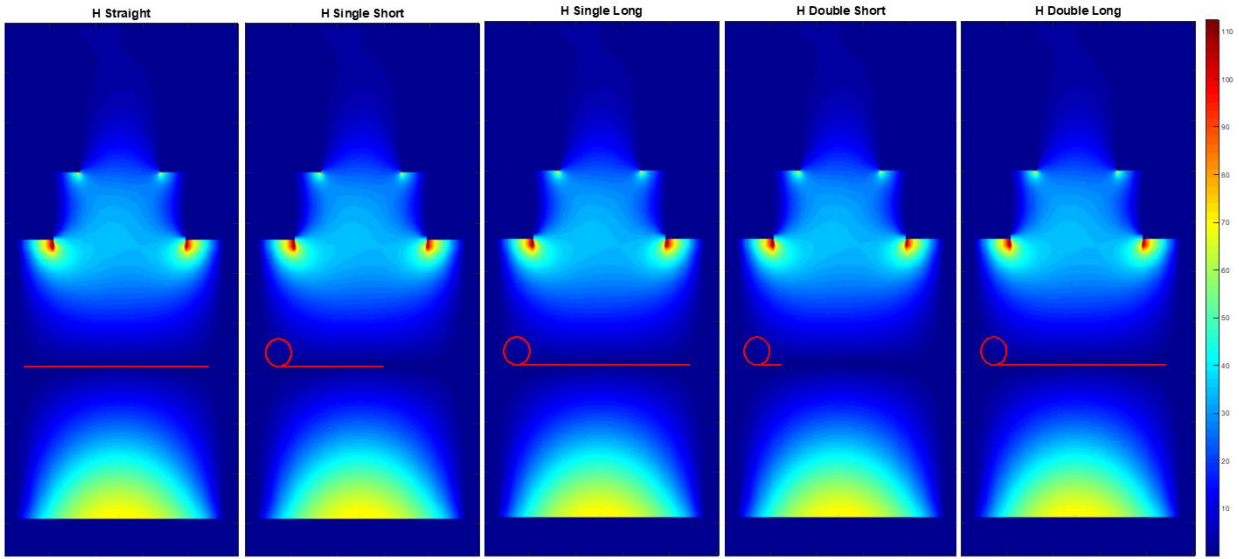

Figure 3: Electric field x-component simulations for the ASTM phantom at $y = 0$ (i.e., a coronal midline slice) at 1.5T showing all the different configurations of the horizontal wires tested. Note that "lead tip" end of the wire corresponds to patient-right end of each horizontally-oriented wire.

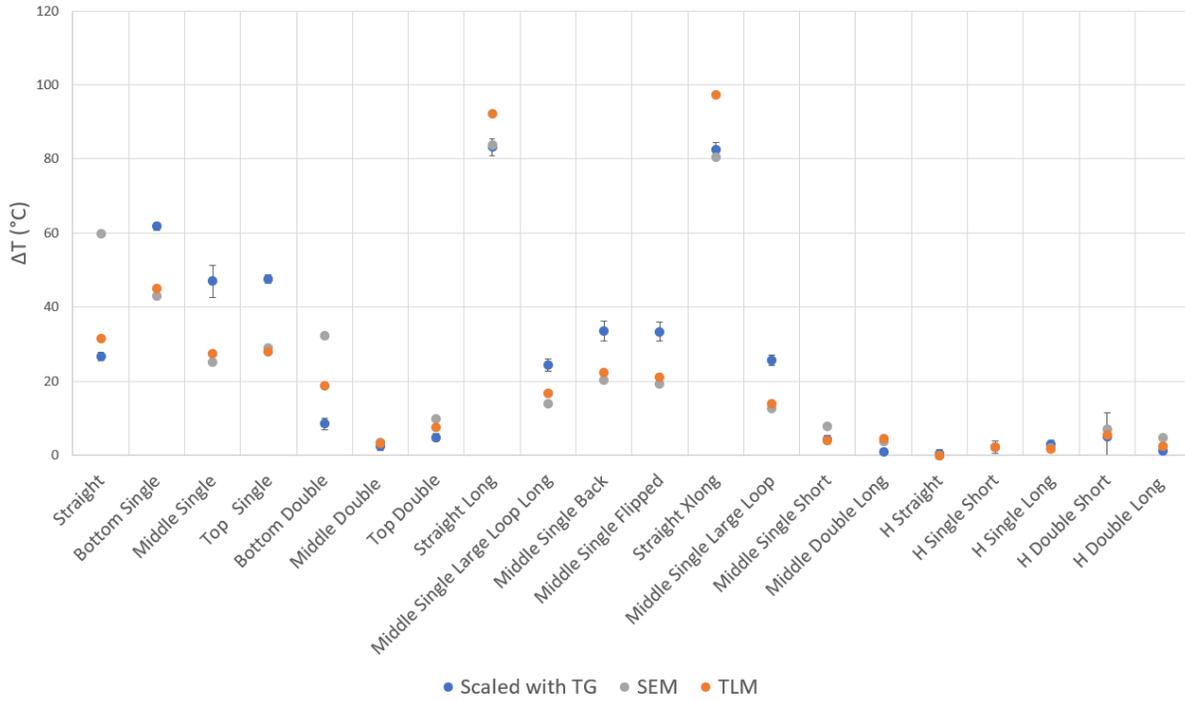

Figure 4: Experimental results of measured temperature rise for all the wires tested as well as the predicted temperature rise $\Delta T$ from the adapted transmission line model (TLM) using an optimized Γ and the simple exponential model (SEM).

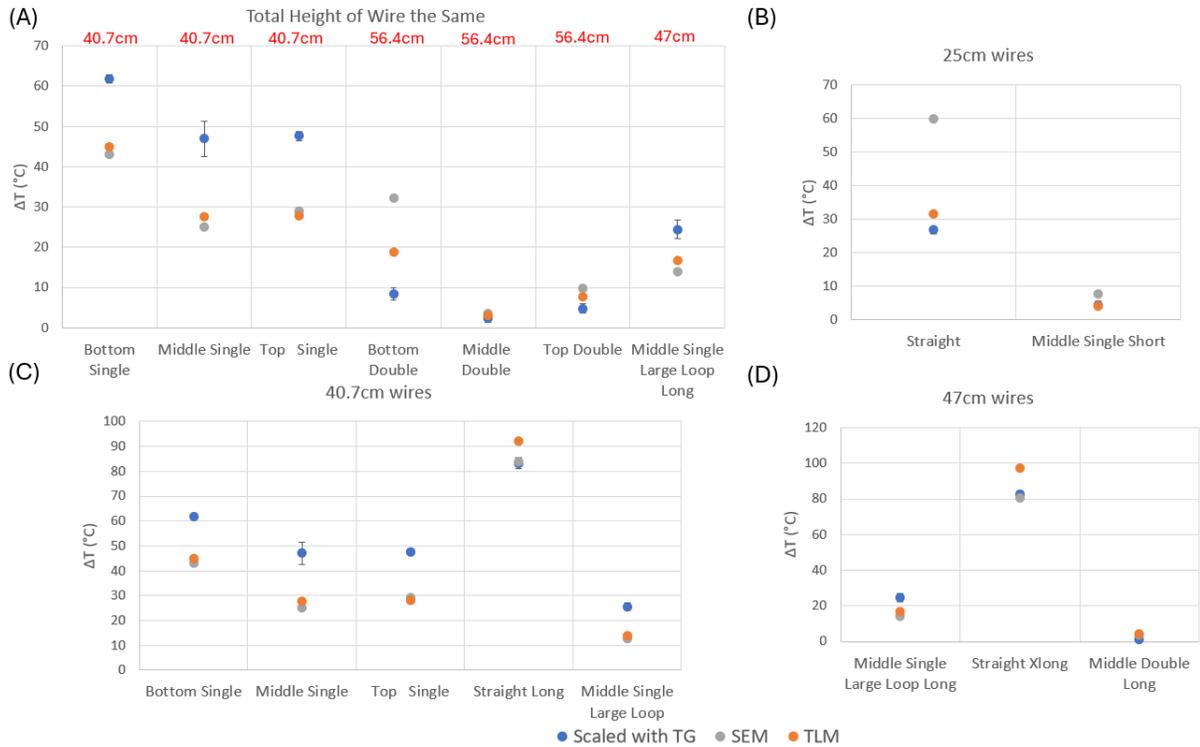

Figure 5: Experimental results for all vertical wires tested as well as the predicted temperature rise $\Delta T$ from both the adapted transmission line model (TLM) using an optimized $\Gamma$ and the simple exponential model (SEM) for various groups including (A) vertical wires with the same total height with the length of each wire shown in red text, (B) vertical wires that were 25cm in length (C) vertical wires that were 40.7cm in length (D) vertical wires that were 47cm in length.

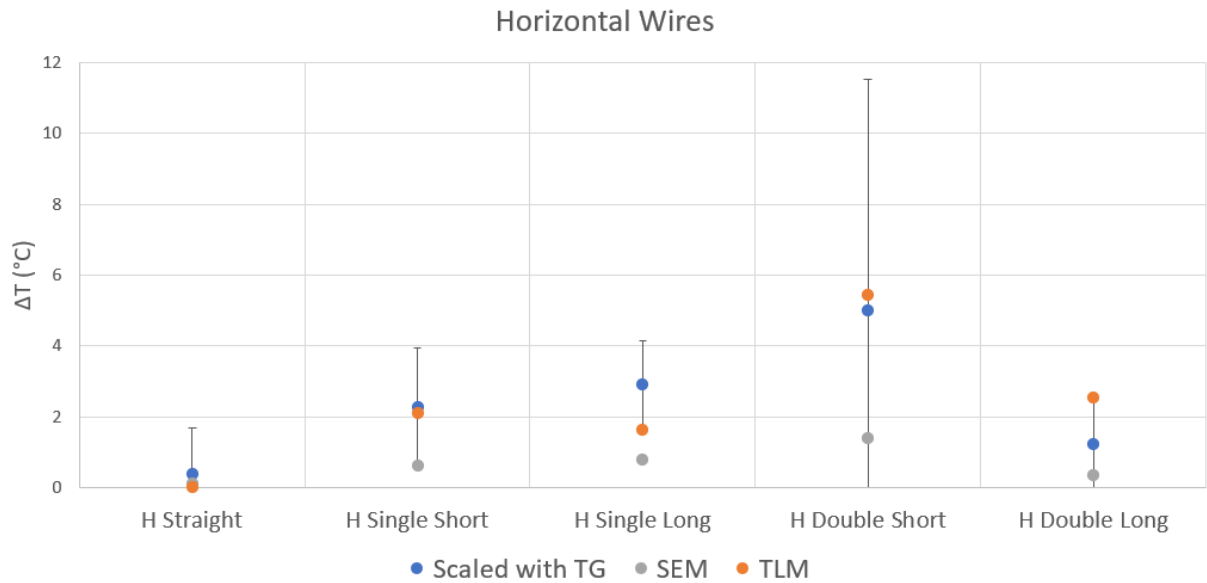

Figure 6: Experimental results for all horizontal wires tested as well as the predicted temperature rise $\Delta T$ from both the adapted transmission line model (TLM) using an optimized $\Gamma$ and the simple exponential model (SEM). Note the range of the temperature rise is quite different from Figure 4.

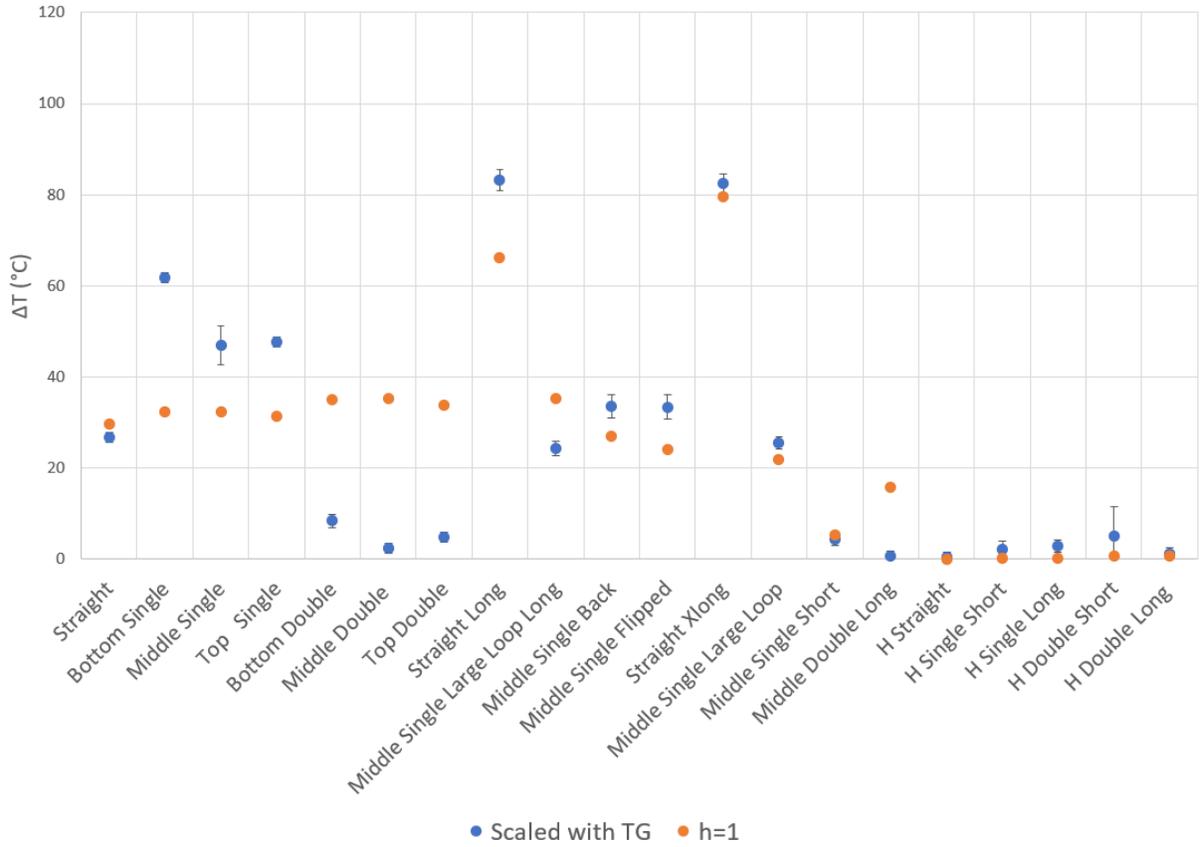

Figure 7: Experimental results of measured temperature rise for all the wires tested as well as the predicted temperature rise $\Delta T$ from $h(l) = 1$.